\documentclass[slac_one]{revtex4}
\usepackage{graphicx}
\usepackage{fancyhdr}
\def\gev{\,{\rm GeV}}

\def\to{\rightarrow}

\def\jp{J/\psi}
\def\be{\begin{equation}}
\def\ee{\end{equation}}
\def\bea{\begin{eqnarray}}
\def\eea{\end{eqnarray}}
\def\bec{\begin{center}}
\def\eec{\end{center}}
\def\atversim#1#2{\lower0.7ex\vbox{\baselineskip\zatskip\lineskip\zatskip
  \lineskiplimit 0pt\ialign{$\matth#1\hfil##\hfil$\crcr#2\crcr\sim\crcr}}}

\begin{document}

\title{Re-visiting Direct $J/\psi$ Production at the
Fermilab Tevatron} %% Paper title goes here


\author{K. Hagiwara}
\affiliation{KEK Theory Division and Sokendai, Tsukuba, Ibaraki
305-0801, Japan}
\author{W. Qi} \affiliation{IHEP, CAS, YuQuan Road 19B, Beijing 100049, China}
\author{C.-F. Qiao} \affiliation{GUCAS, CAS, YuQuan Road 19A, Beijing 100049, China}
\author{J.-X. Wang} \affiliation{IHEP, CAS, YuQuan Road 19B, Beijing 100049, China}

\begin{abstract}
We re-analyze the direct $\jp$ production processes at the Fermilab
Tevatron in view of the recent observation at the B-factories, where
both $\jp$ inclusive and exclusive production rates are found to be
about an order of magnitude larger than the leading order estimates
of non-relativistic QCD. The charm quark fragmentation to $\jp$,
which is the dominant color-singlet process at high $p_T$, is
normalized by the B-factory measurements. The process receives
further enhancement due to the charm sea contribution which has so
far been ignored in most analyses. After summing up all sub-process
contributions, we find that the color-singlet mechanism alone can
account for 20\% to 90\% of the observed direct $\jp$ high $p_T$
production. The polarization rate of the directly produced $\jp$ is
sensitive to the fraction of the color-octet contribution, which is
employed to fill the gap between the color-singlet prediction and
experimental data. With a bigger $K$-factor for the charm quark
fragmentation probability, we envisage a smaller matrix element for
the color-octet $^3S_1^{(8)}$ state, and this can be examined at the
LHC in near future.
\end{abstract}

\maketitle

\thispagestyle{fancy}


\section{INTRODUCTION} % Section title should be in all capitals.

In the history of quarkonium physics, the so-called color-singlet
model (CSM) was widely employed in the study of heavy quarkonium
production and decays \cite{t.a.degrand}. Although CSM made a great
success in the description of quarkonium production and decay in the
past, the Tevatron data \cite{cdf1,cdf3} on large-$p_T$ $J/\psi$
production collected in the 1992-1993 run severely challenged it, as
the data differ very much from the leading order (LO) CSM
predictions in both normalization and $p_T$ distribution.

One of the developments in past decade is the so-called
nonrelativistic quantum chromodynamics (NRQCD) \cite{nrqcd}, which
provides a plausible explanation for the CDF measurements on direct
$J/\psi$ and $\psi^{\prime}$ production at the Fermilab Tevatron
\cite{cdf3}. Within the framework of NRQCD, the surplus charmonia
production in excess of the CSM prediction attributes to the
color-octet contribution. Since the magnitudes of the universal
color-octet matrix elements are determined by fitting to the
experimental data, to find an independent color-octet signature is
necessary to further confirm the applicability of NRQCD to
charmonium production. The color-octet scenario encounters some
difficulties while it is confronted with the new experimental data
\cite{brambilla}. The most striking puzzle probably is the absence
of transversely polarized $J/\psi$ and $\psi'$ in experiment
\cite{wise,kniehl}.

Recently, BaBar and Belle data indicate that the experimental
measurement \cite{babar, belle1, belle2} of the $J/\psi$ inclusive
and exclusive production rates at the B-factories greatly overshoot
the leading-order theoretical calculations in CSM \cite{leibovich}.
Because the $J/\psi$s are found to be produced in accompany with
charmed mesons, very large higher-order corrections to the charm
quark fragmentation process should be inevitable, though
fragmentation mechanism may not work well at or beyond the
next-to-leading order(NLO). According to the analyses in Refs.
\cite{HKLQ,HKQ}, $e^+ + e^- \rightarrow J/\psi + c + \bar{c}$
process requires a $K$-factor of $2.6^{+2.4}_{-1.1}$ to $7.2\pm 2.9$
to mimic the higher order corrections. It was found that with a
constant $K$-factor of around 4 one can reproduce both the inclusive
$J/\psi$ production momentum distribution and the large branching
ratio of $J/\psi +c +\bar{c}$ production observed at the $B$
factories \cite{belle1,belle2,belle3}. Indeed, very recently large
NLO corrections to the exclusive double charmonia production
\cite{chao} and the $\jp + c + \bar{c}$ production processes
\cite{chao1} are obtained. Note that the employed $K$-factor in this
work represents all corrections, rather than merely the NLO QCD one.

In the following, by taking into account the charm-sea induced
processes and a $K$-factor obtained in the $B$-factory in
representing various higher-order corrections to the charm quark
fragmentation mechanism, we re-visit $\jp$ surplus production issue
at the Fermilab Tevatron.

\section{CALCULATION AND RESULTS OF THE $\jp$ HIGH $p_T$ PRODUCTION}

Schematically, the quarkonium hadroproduction process, $A\; +\; B
\rightarrow \psi\; + \; {\sc X}$, can be formulated in a factorized
form,
\bea d\sigma (A &+& B\ \rightarrow\ H(p_T)\; +\; {\sc X}) =
\sum_{a,b} \int_0^1 dx_a \int_0^1 dx_b\; f_{a/A}(x_a,Q) \nonumber \\
&\times& f_{b/B}(x_b,Q)\; d\hat{\sigma}(a + b \rightarrow H(p_T) +
{\sc X})\;, \eea
where $a$ and $b$ are the incident partons within the colliding
hadrons A and B, respectively; $f_{a/A}$ and $f_{b/B}$ are parton
distribution functions at the scale of $Q^2$. In large transverse
momentum, the charmonium hadroproduction are dominated by the
fragmentation sub-processes \cite{braaten1}, which can be expressed
as
\be d\hat{\sigma}(a+b\to H(p_T)+X) = \sum_c\int^1_0 dz\
d\hat{\sigma} (a+b\to c(p_T/z)+X)\ D_{c\to H}(z,Q)\; , \label{eq0}
\ee
where $c$ denotes the fragmenting parton, $D_{c\to H}(z, Q)$ is the
parton fragmentation function. The evolution of the fragmentation
function $D_{c \rightarrow H}(z,\; Q)$ in regard of scale $Q$ in Eq.
(\ref{eq0}) is realized by solving the Altarelli-Parisi (AP).

In our numerical calculation, $J/\psi$ radial wave function at the
origin $R(0)$ is obtained from its leptonic decay width
$\Gamma(J/\psi\to e^+e^-) = 5.55\times 10^{-6}\gev$ \cite{pdg}, and
the CTEQ6L parton distribution functions are employed \cite{cteq}.
To avoid the large logarithms from high order corrections, the
fragmentation, factorization and renormalization scales are set to
be the $J/\psi$ transverse mass, i.e. $m_T = \sqrt{p^2_T +
m^2_{J/\psi}}$.

\begin{figure}
\centering
\includegraphics[width=50mm]{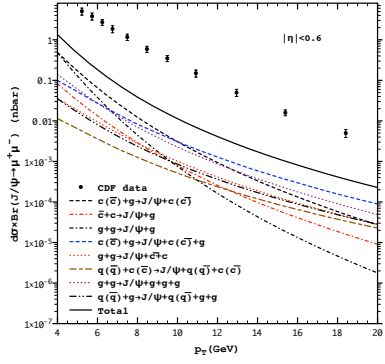}
\caption{Direct $\jp$ production differential cross-sections for
various color-singlet processes versus the Run-I experimental data
from the Fermilab Tevatron.} \label{fig:csptdis}
\end{figure}

Figure \ref{fig:csptdis} shows the transverse momentum distribution
of the directly produced $\jp$ at the Fermilab Tevatron, where eight
main color-singlet sub-processes and their sum are presented
separately. In the literature, the analytic results for most of the
CS processes exist, and we find numerically agreements between our
results and those in previous studies. Different from
Refs.\cite{braaten2,cacciari}, here the gluon fragmentation process
overshoots the charm quark fragmentation process, due to the reason
that the induced charm quark fragmentation in the gluon
fragmentation evolution equation is taken into account in our
calculation. From the figure, one can easily find that two charm-sea
induced sub-processes contribute dominantly to the direct $J/\psi$
production, which in fact partly represent the
next-to-next-to-leading(NNLO) order $g + g \rightarrow J/\psi + X$
process. Although the importance of the second process at very high
transverse momentum was mentioned in Ref. \cite{qiao}, the inclusion
of the first one is necessary at moderate $p_T$ region.

\begin{figure}
\vspace{-.7cm}
 \centering
\includegraphics[width=55mm]{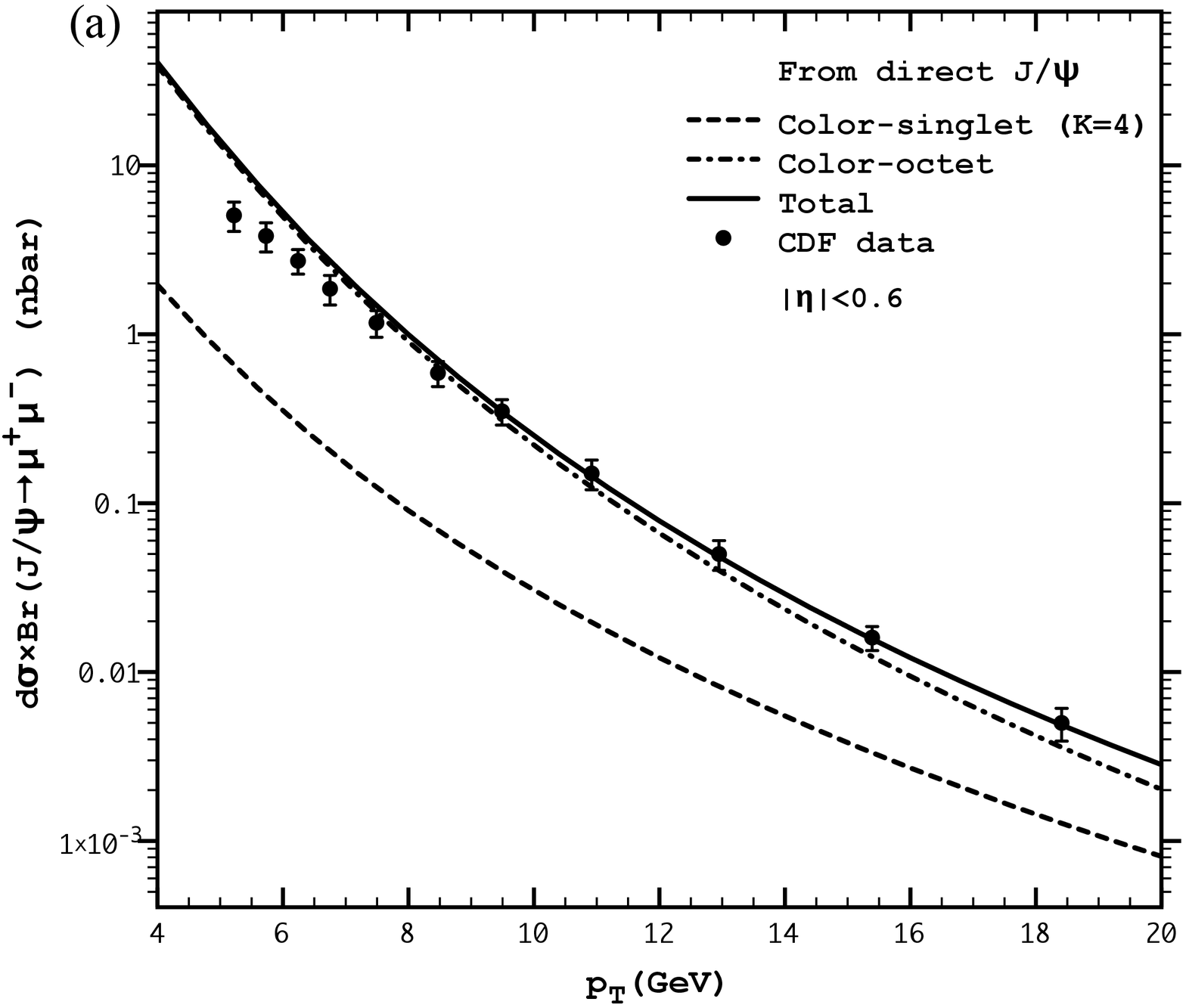}
\includegraphics[width=55mm]{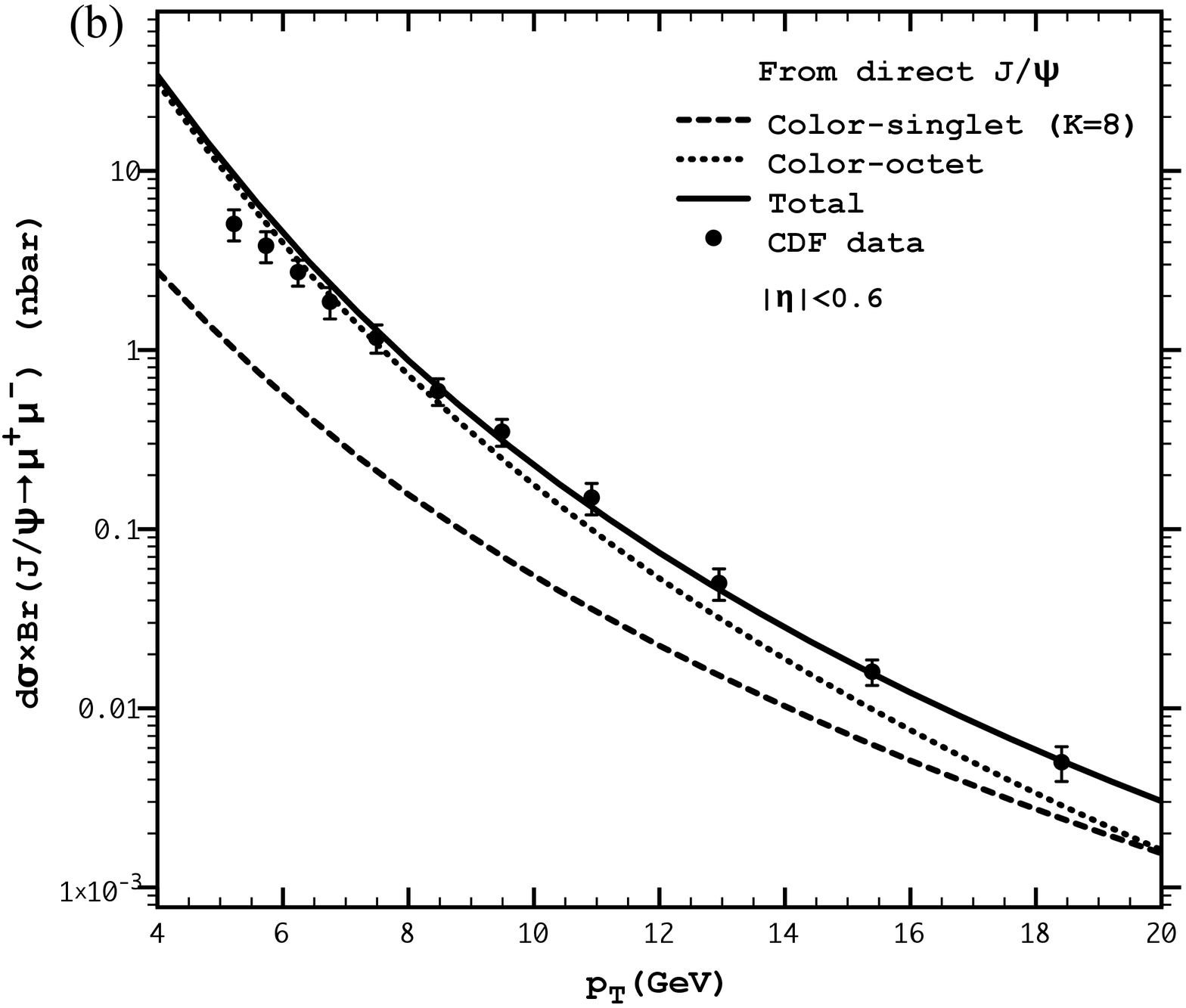}
\includegraphics[width=53mm]{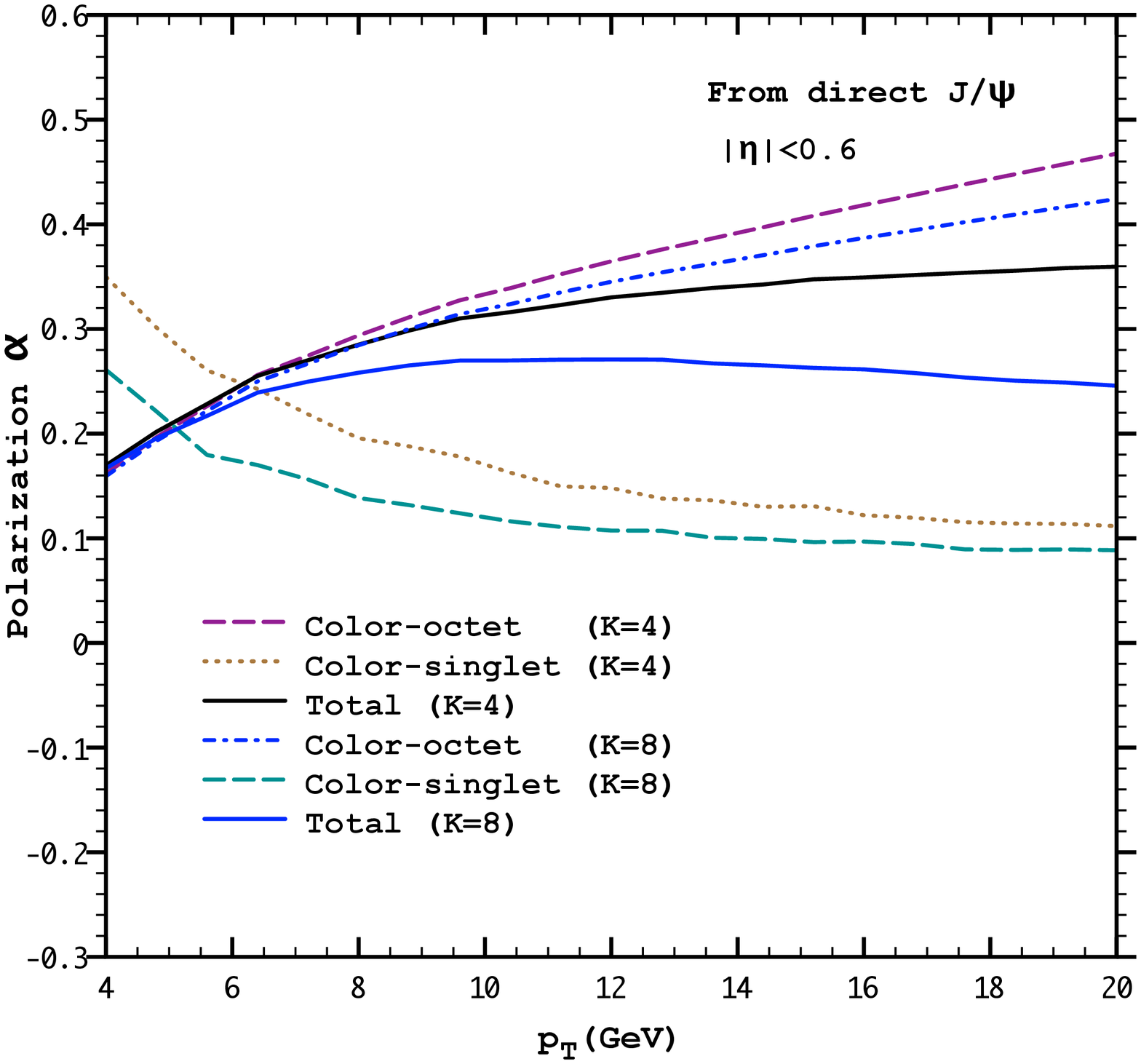}
\caption{NRQCD prediction for the direct $\jp$ production at the
Tevatron Run I energy, where the color-octet matrix element is
adjusted in order to fill the gap between CS prediction and
experimental data. (a) The normalization factor for the charm
fragmentation process is $4\times2$, 4 for $K$-factor and 2 for
hard-scattering cross-section; (b) the normalization factor is taken
to be $8\times2$. (c) The $\jp$ polarization variable $\alpha$
versus transverse momentum.} \label{fig:totptdis}
\end{figure}

Enforcing a factor of 4 to the charm quark fragmentation function
and an overall factor of 2 to the hard scattering processes, we
obtain the re-normalized differential cross sections, as shown in
Fig. \ref{fig:totptdis}(a). Here, the difference between
experimental data and theoretical calculation in CSM reduces to
about a factor of 4. If we assume that the remaining gap attributes
to the color-octet contribution, the matrix element of
$<0|^3\!S_1^{[8]}|0>$ may diminish 1/4 from the initial fitted value
\cite{fleming}. In the case of $K=4$ one cannot find a consistent
explanation for both CDF and Belle results. Therefore, the larger
$K$-factor of $7.2\pm 2.9$, which is the upper limit accommodated by
the Belle data is envisaged. In this case, the color-singlet
mechanism alone explain the data in large transverse momentum
region, as shown in the Fig. \ref{fig:totptdis}(b).

\begin{figure}
\vspace{0.5cm} \hspace{6.5cm}{\tiny{(c)}}
\end{figure}

Figure \ref{fig:totptdis}(c) exhibits our calculation result on
$J/\psi$ polarization distribution versus transverse momentum. Here,
the higher excited state feed down effects are neglected. Of this
figure, the vertical axis denotes the conventional measure of
polarization, defined as: $\alpha = (\sigma_T - 2
\sigma_L)/(\sigma_T + 2 \sigma_L)$; the horizontal axis denotes the
$J/\psi$ transverse momentum. The upper and the lower limits,
$\alpha=1$ and $\alpha=-1$, correspond to the solely transverse and
longitudinal polarization cases respectively, while $\alpha =0$
corresponds to the unpolarized situation. In our result, because the
quark fragmentation process gives unpolarized $\jp$ in the
relativistic limit \cite{ioffi}, the total polarization of directly
produced $\jp$ in high $p_T$ region tends to be unpolarized. This is
completely different from the predictions in color-octet model and
in previous calculations in CSM \cite{braaten2}. \vspace{-1mm}

\section{CONCLUSIONS}

In summary, we have recalculated the direct $J/\psi$ high-$p_T$
production rate in color singlet mechanism and find that the charm
sea induced processes contribute dominantly in the direct $J/\psi$
production while $p_T > 4$ GeV at the Fermilab Tevatron. In recently
the next-to-leading order QCD correction to process (5) is realized
\cite{NQCD,add}. However, as it is well-known that in $J/\psi$ high
$p_T$ hadroproduction the dominant processes are those beyond the
NLO ones. So, only the full NLO calculation is not enough for the
issue discussed in this work.

Based on the $K$-factor scenario we find that the Tevatron data can
be almost accommodated by the color-singlet production mechanism. We
assign $K$-factors of 4 and 8 respectively to the charm quark
fragmentation process, and a factor of 2 to the parton-parton hard
interaction processes. Quantitatively these enhancement factors may
vary from process to process, but for a qualitative analysis we
think a constant number is still meaningful. In the case of
$K$-factor equals to 8, the CDF data can be simply understood in the
color-singlet mechanism.

We have also evaluated the polarization rate of directly produced
charmonium in color-singlet mechanism. We find that at large
transverse momentum, the $\jp$s will be statistically unpolarized.
If this is confirmed by future experiment, then we may conclude that
the gap between our calculation result and experimental data should
be eliminated by giving a bigger $K$-factor to CS processes, but
rather filled by the color-octet component. This is in fact somewhat
favored by the Tevatron II preliminary measurement \cite{cdf5}.
\vspace{-5mm}
\begin{acknowledgments}
This work was supported by the core university program of JSPS. KH
was supported in part by Grant-in-Aid for Scientific Research (No.
17540281) from MEXT, Japan. We thank Z.H. Lin for his initial effort
and contribution to this work. CFQ and JXW were supported in part by
the National Natural Science Foundation of China. CFQ is grateful
for the hospitalities of KEK theory division and ICTP high energy
group for visits while this work was initiated and proceeded.
\end{acknowledgments}
%\vspace{-5mm}

\end{document}